\documentclass[a4paper,twocolumn,superscriptaddress,amsmath]{revtex4-1}
\usepackage{amsmath}
\usepackage{amssymb}
\usepackage{graphicx}
\usepackage[T1]{fontenc}
\usepackage{color}
\usepackage{natbib}

\begin{document}

\title{Magnetic and Electric Mie-Exciton Polaritons in Silicon Nanodisks}
\author{Francesco Todisco}
\email{ft@mci.sdu.dk}
\affiliation{Center for Nano Optics, University of Southern Denmark, Campusvej 55, DK-5230 Odense M, Denmark}
\author{Radu Malureanu}
\affiliation{Department of Photonic Engineering, Technical University of Denmark, DK-2800 Kongens Lyngby, Denmark}
\author{Christian Wolff}
\affiliation{Center for Nano Optics, University of Southern Denmark, Campusvej 55, DK-5230 Odense M, Denmark}
\author{P.~A.~D.~Gon\c{c}alves}
\affiliation{Center for Nano Optics, University of Southern Denmark, Campusvej 55, DK-5230 Odense M, Denmark}
\affiliation{Center for Nanostructured Graphene, Technical University of Denmark, DK-2800 Kongens Lyngby, Denmark}
\affiliation{Department of Photonic Engineering, Technical University of Denmark, DK-2800 Kongens Lyngby, Denmark}
\author{Alexander S. Roberts}
\affiliation{Center for Nano Optics, University of Southern Denmark, Campusvej 55, DK-5230 Odense M, Denmark}
\author{N. Asger Mortensen}
\email{asger@mailaps.org}
\affiliation{Center for Nano Optics, University of Southern Denmark, Campusvej 55, DK-5230 Odense M, Denmark}
\affiliation{Danish Institute for Advanced Study, University of Southern Denmark, Campusvej 55, DK-5230 Odense M, Denmark}
\affiliation{Center for Nanostructured Graphene, Technical University of Denmark, DK-2800 Kongens Lyngby, Denmark}
\author{Christos Tserkezis}
\email{ct@mci.sdu.dk}
\affiliation{Center for Nano Optics, University of Southern Denmark, Campusvej 55, DK-5230 Odense M, Denmark}
\begin{abstract}
     Light-matter interactions at the nanoscale constitute a fundamental ingredient for engineering applications in nanophotonics and quantum optics. To this regard electromagnetic Mie resonances excited in high-refractive index dielectric nanoparticles have recently attracted interest because of their lower losses and better control over the scattering patterns compared to their plasmonic metallic counterparts. The emergence of several resonances in those systems results in an overall high complexity, where the electric and magnetic dipoles have significant overlap in the case of spherical symmetry, thus concealing the contributions of each resonance separately. Here we show, experimentally and theoretically, the emergence of strong light-matter coupling between the magnetic and electric-dipole resonances of individual silicon nanodisks coupled to a J-aggregated organic semiconductor resonating at optical frequencies, evidencing how the different properties of the two resonances results in two different coupling strengths. The energy splittings observed are of the same order of magnitude as in similar plasmonic systems, thus confirming dielectric nanoparticles as promising alternatives for localized strong coupling studies. The coupling of both the electric and magnetic dipole resonances can offer interesting possibilities for the control of directional light scattering in the strong-coupling regime and the dynamic tuning of nanoscale light-matter coupled states by external fields.
\end{abstract}
\maketitle

\section{Introduction}
Light-matter interactions constitute a fundamental field of study in photonics, since it opens routes for exploring novel physical effects and exploiting applications in optoelectronics and quantum optics.\cite{Walls} Although this field started growing around high-quality-factor cavities with extremely low losses and diffraction limited mode volumes,\cite{Raimond2001,Mabuchi2002} in recent years increasing interest has been devoted to push the research down to the nanoscale by using open nanocavities, with the promise of shrinking light-matter interaction lengths down to scales much smaller than a single wavelength\cite{Hummer2013} and manipulating the generation of nonclassical light by quantum emitters.\cite{Fernandez-Dominquez2018} 

To this regard, plasmonic resonances localized in metallic nanoparticles have been so far the preferential studied platform. Indeed, their ability to enhance and concentrate light in extremely sub-wavelength volumes, together with their high sensitivity to the environment's refractive index, make them ideal to study the interactions with nearby molecules or quantum emitters. As a result, exciting effects have been observed so far, including Purcell enhancement,\cite{Vesseur2010,askelrod2014} enhanced Raman\cite{Jackson2004} and fluorescence\cite{Chance1978,Fu2009} spectroscopy, localized heating\cite{baffou2013} and sensing.\cite{Anker2008,galush2009}

Of particular interest is the case when light-matter interactions enter the so-called strong-coupling regime. This happens when the coupling strength exceeds the overall system losses, meaning that (nano)cavity photons and nearby emitters can coherently exchange energy, while the emitter occupation oscillates at the so-called Rabi frequency. As a result the system becomes an effective mixture of light and matter components, named polaritons, characterized by a peak splitting in the scattering and absorption spectrum and an anti-crossing in the dispersion. More importantly, as effective light-matter mixtures, polaritons offer the unique potential of dressing photons with nonlinearities.\cite{delteil2019,Verger2006}

In the field of plasmonics, strong coupling has been widely investigated when integrating metallic nanoparticles with organic and inorganic materials sustaining excitonic transitions.\cite{torma} Under the proper conditions, J-aggregating molecules,\cite{Bellessa2012,Todisco2015,zengin2013} dyes,\cite{Todisco2018} quantum dots\cite{gomez2010} and two-dimensional (2D) materials\cite{stuhrenberg2018,Geisler2019} placed nearby resonant metallic nanoparticles or nanoparticle arrays, have been shown to give rise to strong plasmon-exciton coupling. More importantly, many interesting effects have been predicted and observed in these systems, including coherent emission,\cite{Bellessa2012} lasing in nanoparticle arrays,\cite{Ramezani:17,DeGiorgi2018,hakala2018} ultrafast Rabi oscillations,\cite{Vasa2013} chemical dynamics tuning,\cite{Munkhbat2018} coupling with dark excitonic states\cite{cuartero2018} and few- to single-emitter coupling.\cite{chikkaraddy2016,santhosh2016} However, high absorption losses in metallic materials still remain the main drawback of studying strong-coupling effects in plasmonics, given the resulting broad resonances with extremely short lifetimes, of the order of few femtoseconds. Moreover, plasmonic materials suffer from heating effects that can hinder the access to nonlinearities, and from a poor integration with existing device technology,\cite{Naik2013} thus encouraging the explorations of alternative platforms for studying light-matter interactions at the nanoscale.

From this point of view, high-refractive-index dielectrics have recently been considered as a possible alternative for generating localized optical resonances at the nanometer length scale.\cite{Kuznetsov2016} In this case, in fact, rather than from collective oscillations of free charge carriers, optical resonances arise as a result of the oscillation of polarization charges and the circular displacement current inside the particles when the light wavelength and the particle size become comparable.\cite{YANG2017} Even though this behaviour is well known since Mie's works on scattering by small particles,\cite{Bohren1983} it is now raising new interest because of the unique possibilities offered in terms of the excitation of magnetic dipolar and multipolar resonances, the control of the light scattering pattern,\cite{staude2013,Cihan2018} and the presence of anapole resonances.\cite{yuanqing2018} The optical properties of these particles have been investigated for applications in sensing,\cite{quidant2017} metasurfaces,\cite{Yu2015} local field enhancement and nonlinearities.\cite{Gigli:19} Furthermore strong light-matter coupling has been predicted and investigated in these systems when interacting with excitonic transitions, and it was recently shown in 2D transition metal dichalcogenides nanodisks, simultaneously hosting the excitonic and the Mie resonances.\cite{Verre2019} On the other hand, high-refractive-index nanospheres have been theoretically studied showing the accessibility of the strong-coupling regime with dyes having a sufficient oscillator strength,\cite{Tserkezis2018} while experiments in a similar system have shown a peak splitting appearing well below the strong-coupling criterion,\cite{Wang2016} thus being only an induced-transparency effect.\cite{antosiewicz2014} Moderate-refractive-index nanoparticles have been investigated as well, showing a weak-coupling behaviour.\cite{Ruan2018}

Here we report on the onset of the strong-coupling regime between the Mie resonances of silicon nanodisks and a molecular J-aggregate. Unlike nanospheres, lower symmetry geometries enable to independently tune the magnetic-dipole (MD) and electric-dipole (ED) resonances.\cite{vandeGroep2013} In this way we studied the light-matter coupling behaviour of the two separately, finding both the resonances to be above the onset of the strong-coupling regime. Experimental and numerical results show that the MD of small nanodisks is characterized by a larger coupling strength with respect to the ED of larger nanodisks resonating at the same frequency. This behaviour can be understood in terms of the different near-field profiles of the two resonances, resulting in a different overlap with the thin excitonic layer.

\section{Results and Discussion}\label{Sec:discussion}
Our sample consists of silicon nanodisks (NDs) on a sapphire substrate. Sapphire has a slightly higher refractive index ($n=1.775$ at $\lambda=500$\,nm) than other commonly used substrates like quartz or glass, but this has been shown to result in only a slight increase of the radiative damping and a broadening of the MD and ED resonances of the supported dielectric particles, especially in the cases when a large contact area exists between nanoparticles and substrate.\cite{vandeGroep2013} The cylindrical geometry, as we will show in the following, was chosen because of the straightforward fabrication procedure, and the possibility to independently tune the MD and ED resonances by geometric considerations, a possibility that is not offered in the more symmetric case of nanospheres.\cite{evlyukhin2012}

\begin{figure*}[t]
\includegraphics[width=0.8\textwidth]{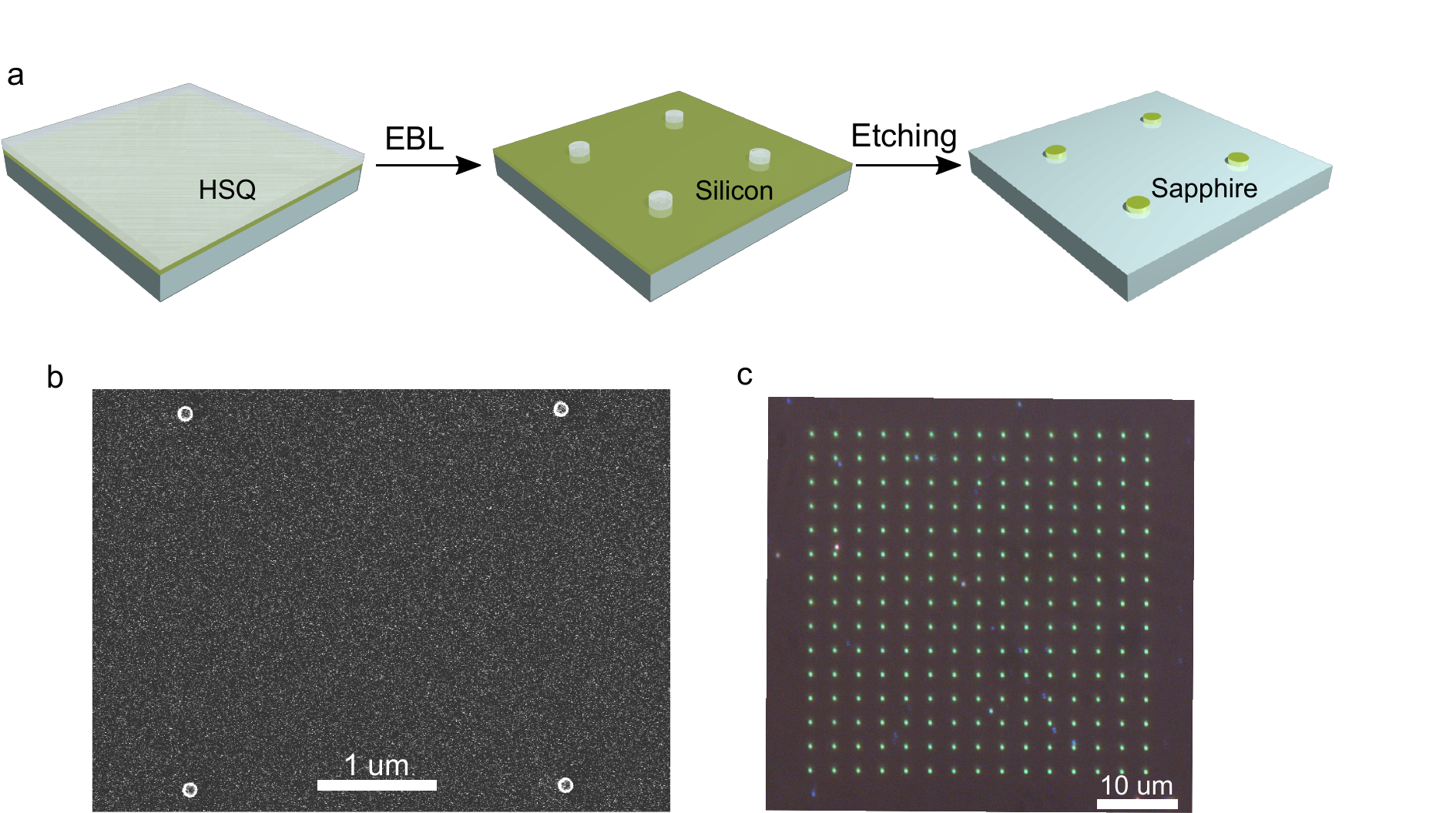}
\caption{(a) Sketch of the nanofabrication procedure. Hydrogen silsesquioxane was used as an electron-beam resist for electron-beam lithography on a silicon on sapphire substrate, followed by plasma etching of silicon and wet etching of the exposed resist. (b) Scanning electron micrograph and (c) dark field optical image of typical fabricated structures arranged in a 3\,$\mu$m pitch, with 110\,nm and 150\,nm diameter nanodisks, respectively.}
\label{fig:fabrication}
\end{figure*}

The nanofabrication approach is sketched in Figure~\ref{fig:fabrication}a. A silicon-on-sapphire (SOS) substrate with a $100\pm10$\,nm thick silicon layer was spin coated with a 200\,nm thick hydrogen silsesquioxane (HSQ), used as a negative-tone resist for electron-beam lithography. NDs with different diameters (ranging from 100\,nm to 300\,nm) were fabricated, arranged in 15$\times$15 square lattices with 3\,$\mu$m pitch, so as to minimize near-field interactions between neighboring particles and enable single-particle measurements. The sample was then plasma etched, followed by BHF wet etching to remove the remaining exposed resist (see Methods section for the details). Scanning-electron microscope and optical dark-field microscope images of the resulting final sample are shown in Figure~\ref{fig:fabrication}b--c, respectively.

The sample was optically investigated by dark-field spectroscopy on an inverted optical microscope. Broadband light from a tungsten lamp was focused at a grazing angle of $\theta\sim 78^\circ$ by a 100$\times$, 0.95 numerical aperture objective lens, used also to collect the back-scattered light. The real image of the sample was then reconstructed on the entrance slits of an imaging spectrometer equipped with a CCD camera. In this way, the proper in-plane alignment of the sample enables the simultaneous measurement of all the aligned 15 nanoparticles in a lattice column (see Supporting Information).

\begin{figure*}[t!]
\centering
\includegraphics[width=0.9\textwidth]{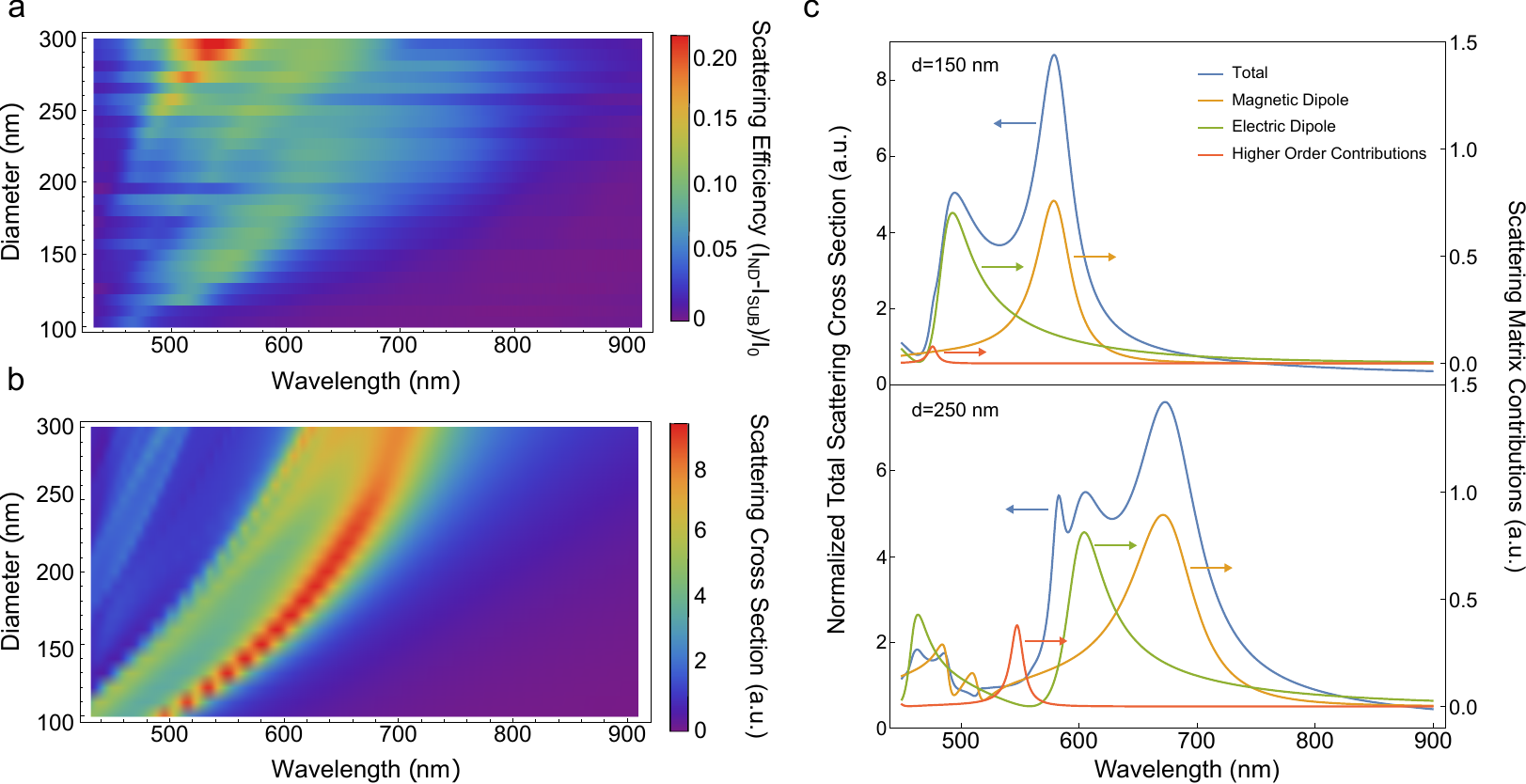}
\caption{Experimental (a) and simulated (b) scattering spectrum dispersion of individual silicon nanodisks, shown as a function of the particles' diameter. The visible Mie resonances show a continuous redshift with increasing nanoparticle size, while higher-order contributions appear for larger particles. (c) Calculated resonance decomposition for small (150\,nm) and large-diameter (250\,nm) nanodisks. The first observable resonance for small particles is a magnetic dipole. The second peak, appearing at higher energy for larger particles is instead an electric dipole, while multipolar contributions also appear at even shorter wavelengths.}
\label{fig:bareSP}
\end{figure*}

The measured ND back-scattering spectra in the investigated diameter range are shown in Figure~\ref{fig:bareSP}a. Here, bright optical resonances appear as scattering peaks, spanning the whole visible to near-infrared spectral range, and redshifting with increasing ND diameter. In particular, small diameter NDs are characterized by an individual sharp resonance at short wavelengths, resulting in bright colors when observed under dark-field excitation (Figure~ \ref{fig:fabrication}c). As the ND diameter exceeds 150\,nm this resonance becomes weaker, while higher-order resonances of comparable strength appear at shorter wavelengths, continuously redshifting as the ND diameter increases. This behaviour, that is well-known in high-refractive-index NDs,\cite{decker2016} can be exploited to achieve highly directional scattering when superimposing the MD and ED resonances.\cite{staude2013,Cihan2018}

To shed light on the nature of the observed resonances, we explored the system electrodynamics with the aid of the extended boundary-condition method (EBCM) for the evaluation of the scattering $T$ matrix of
nonspherical particles.\cite{Mishchenko2002} Since EBCM is based on expanding the incident and scattered fields in spherical waves, the method provides a straightforward way to characterise the resonances in terms of angular-momentum indices, thus assigning a predominant electric (magnetic) dipole, quadrupole etc. character to them (see Methods). In the case of spherical particles, the
method reduces to the well-known Mie theory.~\cite{Bohren1983}

The results are shown in Figure~\ref{fig:bareSP}b, resembling the experimentally observed dispersion. The scattering spectra of two representative diameter NDs (in air) were further analyzed and decomposed in the contributions of the different elementary resonances through the corresponding elements of the $T$ matrix (see Methods), as shown in Figure~\ref{fig:bareSP}c. Here the first peak appearing in small-diameter NDs ($d=150$\,nm) can be ascribed to a MD resonance. According to Mie theory for spherical nanoparticles, this resonance arises when the effective wavelength of light inside the particle ($\lambda/n$, with $n$ being the refractive index of the particle) equals its diameter $d$. This is also the case here for the smallest NDs (where the diameter and the height of the ND are both 100\,nm, and it is reasonable to approximate the particle shape by that of a sphere), considering that the refractive index of silicon at $\lambda=470$\,nm is approximately $n\sim 4.5$\cite{aspnes1983} and the observed peak is centered at $\lambda=470$\,nm. As observed experimentally, the MD redshifts for larger NDs, while resonances predominantly arising from an ED and higher order multipoles appear at shorter wavelengths (see Figure~\ref{fig:bareSP}c).

\begin{figure}[t!]
\includegraphics[width=0.9\columnwidth]{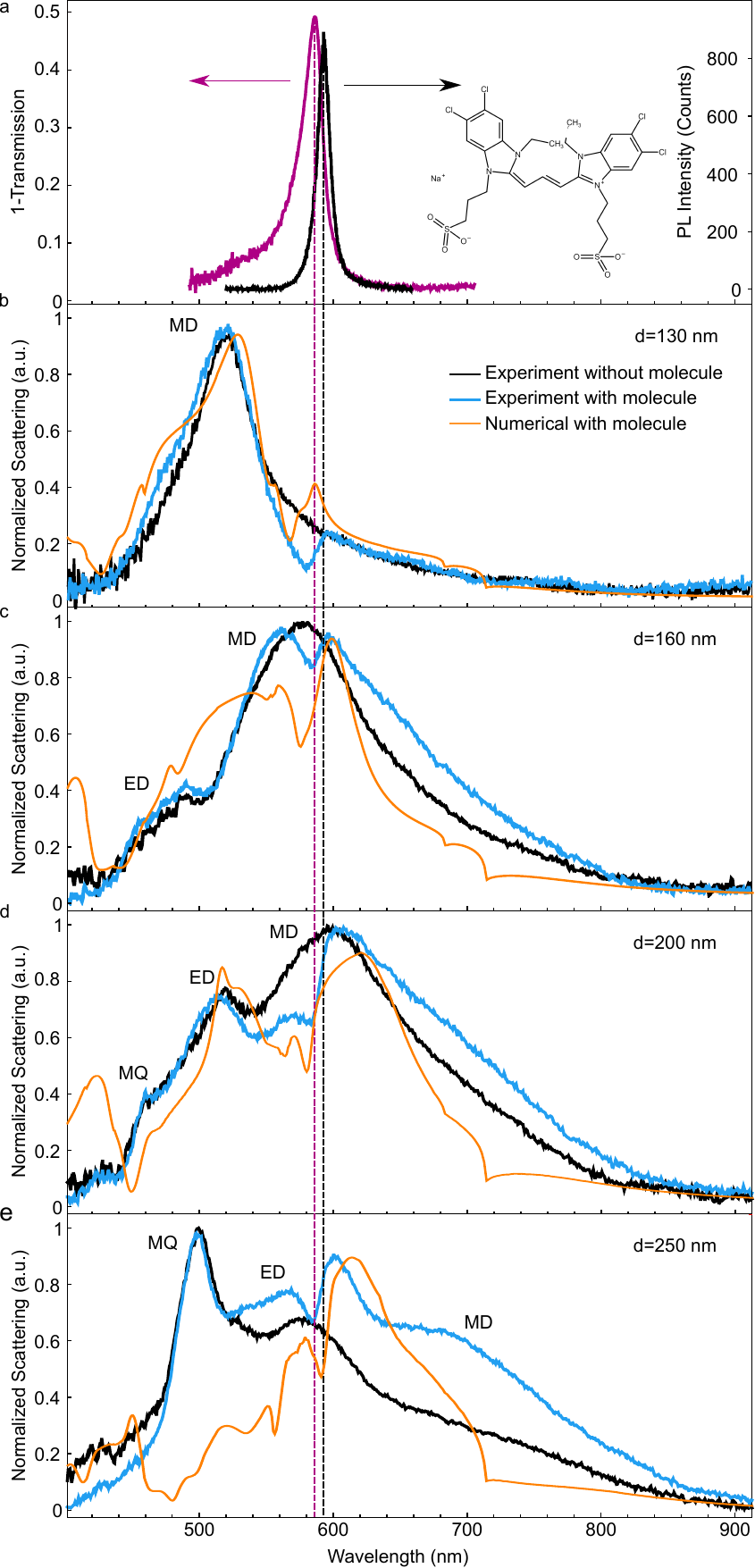}
\caption{(a) Absorption (red) and photoluminescence (black) spectra of the J-aggregated dye (TDBC), whose chemical structure is shown in the inset. (b--e) Evolution of the Mie-exciton coupling as a function of the ND diameter. The black curves correspond to the nanodisks spectra without molecule, while the blue curves are the corresponding spectra in presence of the dye. The orange spectra are numerically calculated spectra for a silicon ND with a thin (20\,nm) dye layer on top (see Methods).}
\label{fig:molecule}
\end{figure}

In order to investigate light-matter interactions supported by silicon NDs, an organic J-aggregating cyanine dye (5,6-dichloro-1-ethyl-3-sulfobutyl-2-(3-(5,6-dichloro-1-ethyl-3-sulfo\-butyl benzimidazolinylidene-1-propenyl) benzimidazolium hydroxide, inner salt, sodium salt or TDBC) was drop-casted from a dilute methanol solution on the sample (see Methods for the details). This particular dye was chosen because of its high-energy J-aggregate absorption band centered at 585\,nm (shown in Figure~\ref{fig:molecule}a), where the MD and ED resonances of the silicon NDs are sharp and well separated in energy (see Figure~\ref{fig:bareSP}a), so as to enable the investigation of the coupling behaviour of the two resonances separately. The sample was then studied by dark-field back-scattering spectroscopy, whose results are shown in Figure~\ref{fig:molecule}b--e and compared to numerical results (see Methods for details). For small diameter NDs, whose MD resonance is highly detuned with respect to the J-aggregate exciton energy (Figure~\ref{fig:molecule}b), the scattering spectrum is characterized by an intense peak at the MD resonance, whose energy position is unaltered with respect to the case without molecule. A lower-intensity asymmetric peak also appears around the exciton energy, that can be ascribed to photoluminescence (PL) from the J-aggregated dye, shown as a black line in Figure~\ref{fig:molecule}a (see Methods for details). On the other hand, when the MD is resonant at the exciton energy (Figure~\ref{fig:molecule}c--d) a peak splitting arises, while the ED peak remains unchanged. The same behaviour is observed as the diameter of the NDs is further increased (Figure~\ref{fig:molecule}e): here the ED is tuned to be resonant with the exciton energy, while the MD resonance appears as a broad peak at longer wavelengths. This behaviour suggests that, in our geometry, no significant cross-talk takes place between the MD and the ED, and that the two resonances couple with the exciton mode without affecting each other. We note here that, although a further increase of the ND diameter would also result in the resonant tuning of higher order resonances at the exciton energy, in this case the presence of several contributions centered at similar energy would not enable a clear analysis of the coupling (see Supporting Information).

\begin{figure}[t!]
\includegraphics[width=0.9\columnwidth]{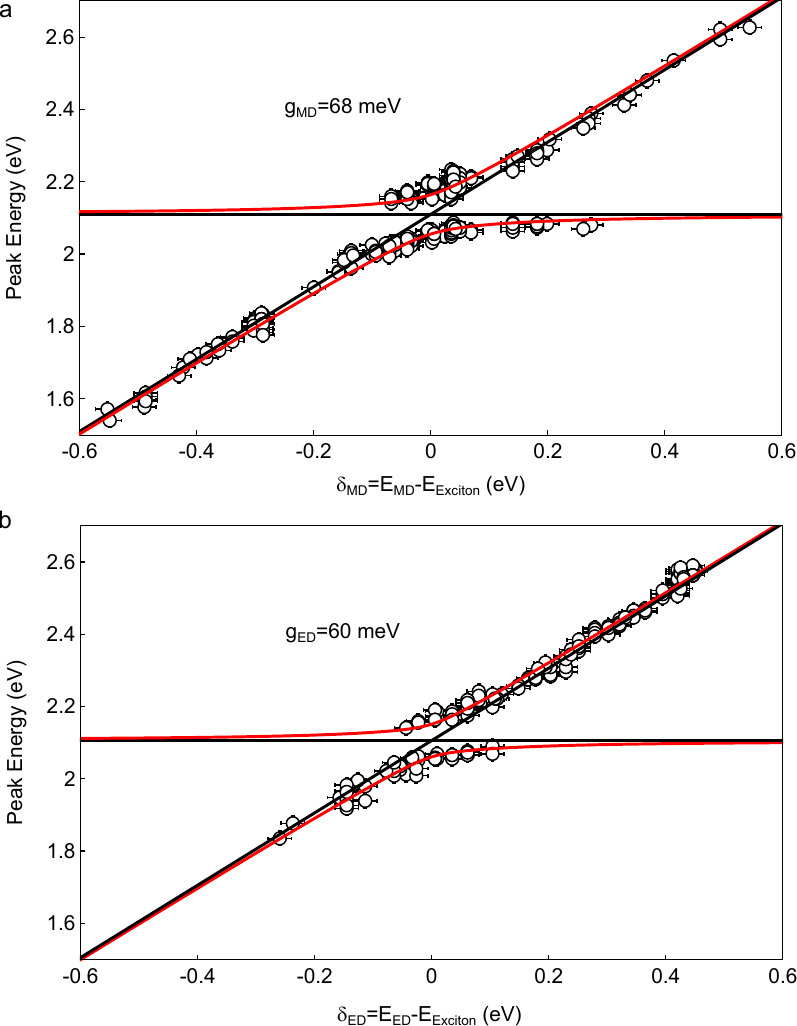}
\caption{Anti-crossing behaviour of the observed peak splitting as a function of the bare Mie-exciton detuning for the magnetic dipole (a) and the electric dipole (b) resonances. Red lines show the fitting based on a two coupled oscillators model, with only the coupling strength $g$ as a fitting parameter. The fitting result is indicated in the inset.}
\label{fig:anticrossing}
\end{figure}

To verify the nature of the Mie-exciton light-matter coupling for the MD and the ED, we extracted the peaks' energy positions for all the fabricated diameters, as shown in Figure~\ref{fig:anticrossing} as a function of each resonance (MD/ED) detuning $\delta$ with respect to the exciton energy. Here an anti-crossing behaviour arises when both the MD and ED cross the exciton energy ($E_{\rm exc}=\hbar\omega_{\rm exc}=2.11$\,eV). Considering that the MD and ED are mutually non-interacting, each resonance dispersion can be fitted independently with a standard two-coupled-oscillators model:
\begin{equation}
  \left(
\begin{array}{ccc}
\omega_1-i\frac{\gamma_1}{2} & g \\
g & \omega_2-i\frac{\gamma_2}{2} \\
\end{array} \right)\left(
\begin{array}{ccc}
a_{1} \\
a_{2} \\
\end{array} \right)=\omega\left(
\begin{array}{ccc}
a_{1} \\
a_{2} \\
\end{array} \right)
\end{equation}
where $\omega_j$ ($j = 1,2$) are the energies of the uncoupled initial states $a_{j}$, and $g$ is the coupling strength between the selected Mie resonance ($j=1$) and the excitonic transition of the dye ($j=2$). In our case $\gamma_2=\gamma_{\rm exc}=40$\,meV, while $\hbar \gamma_1=\hbar \gamma_{\rm MD}\simeq120$\,meV for the MD and $\hbar \gamma_1=\hbar \gamma_{\rm ED}\simeq150$\,meV for the ED resonances, as obtained by fitting the spectra (see Supporting Information). We can thus fit the experimental data with only the coupling constant as a fitting parameter, resulting in a coupling strength $\hbar g_{\rm MD}=68\pm2$\,meV for the MD resonance and $\hbar g_{\rm ED}=60\pm3$\,meV for the ED resonance. A similar result was found by numerical calculations on the coupled system (see Methods for details), obtaining $\hbar g_{\rm MD}=72\pm2$\,meV and $\hbar g_{\rm MD}=59\pm2$\,meV.

The condition to formally evaluate if one system is in the strong-coupling regime has been recently discussed in the literature.\cite{torma,Khitrova} Briefly, the solution of the two-coupled-oscillators problem at zero detuning ($\omega_2=\omega_1=\omega_0$) reads
\begin{equation}
\omega_{\pm}=\omega_{0}-i\frac{\gamma_1}{4}-i\frac{\gamma_2}{4}\pm\frac{1}{4}\sqrt{\left(4g\right)^{2}-\left(\gamma_1-\gamma_2\right)^2},
\end{equation}
where $\omega_{\pm}$ are the energies of the of the resulting hybrid states.
It may then seem that a condition for defining the onset of the strong coupling could rely on the maximization of the splitting, maximizing the term in the square root, that is $2g>\left | \gamma_{1}-\gamma_{2}\right |/2 $. However, as we will explain in the following, in the presence of significative damping this condition can be interpreted as only necessary but not sufficient. 

To ease our discussions, we define the following quantities: $2\omega_{12} \equiv \sqrt{(4g)^2-(\gamma_1-\gamma_2)^2}$ and $ Q_{12}\equiv  2\omega_{12}/(\gamma_1+\gamma_2) $. Next, solving for the dynamics of an initially excited emitter and an empty cavity, one can obtain exact analytical expressions for the occupation factors for the emitter and the cavity, i.e. $\left|a_1(t)\right|^2$ and $\left|a_2(t)\right|^2$. Whereas the literature offers several more or less appealing figures of merit, the point here is that the above parameterizations constitute the one and only natural re-scaling of the time variable in the context of the analytically exact solution to the dynamics (see Supporting Information). In the large-$g$ limit, the solutions are straightforwardly
\begin{equation}
\left\{\begin{matrix}\left|a_1(t)\right|^2\\
\left|a_2(t)\right|^2\end{matrix}\right\}
\simeq 
\exp \left(-\frac{\omega_{12} t}{Q_{12}}\right)\times
\left\{\begin{matrix}
\sin^2\left(\tfrac{1}{2}\omega_{12} t\right)\\\cos^2\left(\tfrac{1}{2}\omega_{12} t\right)\end{matrix}\right\}.
\end{equation}
Thus, $\omega_{12}$ is indeed the characteristic Rabi-like frequency of the oscillations, while in the spirit of cavity ring-down-spectroscopy $Q_{12}$ is a quality factor that quantifies the number of oscillations of the system, before the overall occupation decays in the presence of damping, i.e., $\left|a_1(t)\right|^2+\left|a_2(t)\right|^2 \simeq \exp \left(-\frac{\omega_{12} t}{Q_{12}}\right)$. Similar figures of merit expressed in a slightly different notation have appeared elsewhere~\cite{yang2016}. As such, $Q_{12}=1$ unambiguously marks the transition from weak to strong coupling. In other words, strong coupling implies that $2g >  2g^*\equiv \sqrt{\frac{\gamma_1^2+\gamma_2^2}{2}}$. This inequality, in the limit of $\gamma_1\sim\gamma_2$ becomes $2g \gtrsim \frac{\gamma_1+\gamma_2}{2}$, a condition that was shown to coincide with the coupled-system spectrum becoming flat-topped.\cite{Khitrova} Both these inequalities are usually used in the literature for defining the threshold for the strong-coupling regime. In our case, considering that $\hbar \gamma_{\rm exc}=40$\,meV, $\hbar\gamma_{\rm MD}\simeq120$\,meV and $\hbar\gamma_{\rm ED}\simeq150$\,meV, we get $2 \hbar g^*_{\rm MD}\sim 90$\,meV and $2 \hbar g^*_{\rm ED}\sim 110$\,meV thus demonstrating that both the MD and the ED enter the strong-coupling regime. These splittings are nevertheless of the same order of magnitude as those predicted for silicon nanospheres in a core-shell geometry,\cite{Tserkezis2018,alu2018} and as the ones reported for plasmonic systems on individual and arrays of particles interacting with the same molecule,\cite{Todisco2015,zengin2013} thus confirming that high-refractive-index nanoparticles can be a valid alternative platform to metals, for strong-coupling applications.

\begin{figure}[t!]
\includegraphics[width=0.9\columnwidth]{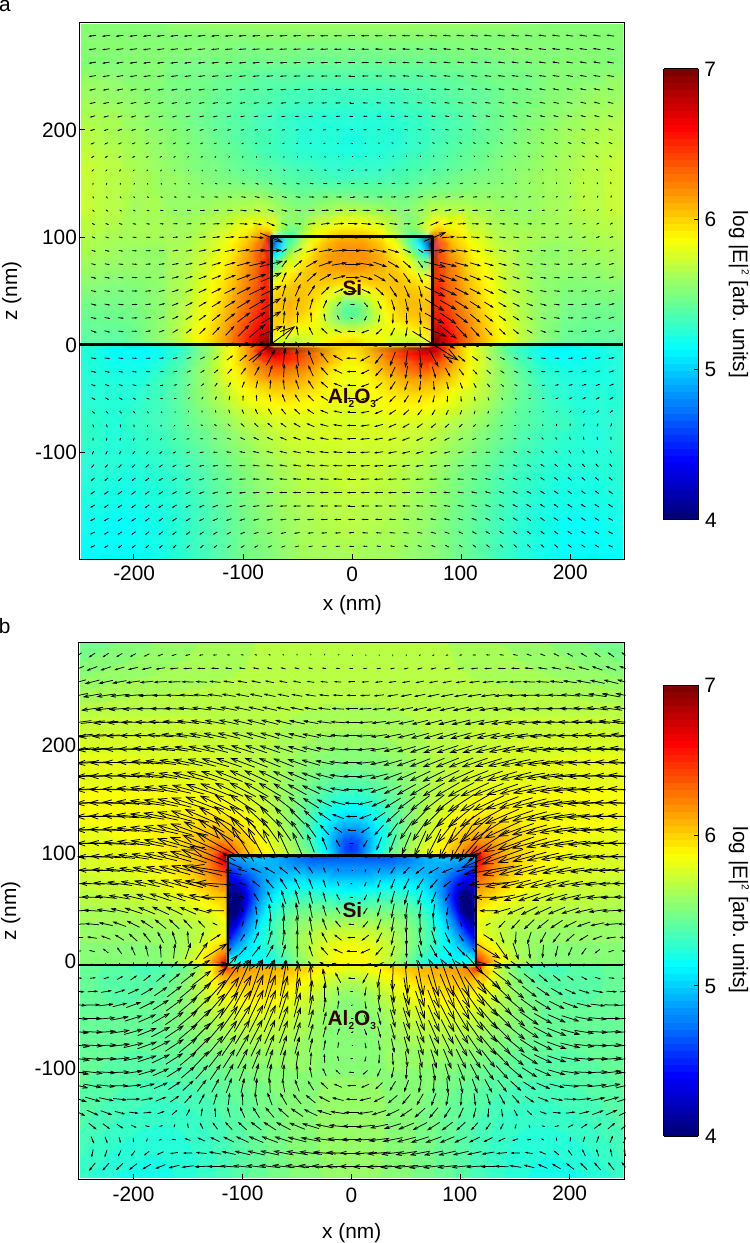}
\caption{Logarithm of total electric field intensity (colour scale) and orientation of the real in-plane part of the electric field (arrows) at resonance and under normal incidence, calculated for silicon nanodisks with (a) 150\,nm and (b) 230\,nm diameter. Here $z<0$ represents the sapphire substrate. The smaller disk clearly exhibits a magnetic-dipole-like behaviour, with a loop in the electric field amplitude inside the particle, while the larger disk shows an electric-dipole-like behaviour, with an antinode at the nanodisk center.}
\label{fig:maps}
\end{figure}

Another interesting effect observed in our system is the difference in the coupling strength between the MD and the ED resonances, observed both experimentally and numerically. Contrary to what one might perhaps presume, the MD shows a larger coupling strength compared to the ED. In general, since the electromagnetic field profiles are expected to be spatially different for the two resonances, there is every reason to expect the frequency splittings to differ as well. This is because the coupling constant $g$ can be considered as an indirect measure of the overlap between the electric field of one optical mode and the dye in the near-field of the nanoparticle. 
In order to check this, we performed numerical calculations on a silicon ND geometry excited at normal incidence with a plane wave, and calculated the electromagnetic field total intensity (colour scale) and the electric field amplitude (arrows) at resonance, in a vertical plane cutting the silicon ND in its centre, as shown in Figure~\ref{fig:maps}. We chose the two ND diameters corresponding to the zero detuning condition for the MD ($d=150$\,nm, Figure~\ref{fig:maps}a) and the ED ($d=230$\,nm, Figure~\ref{fig:maps}b); for a short discussion on the differences in sizes between experiments and simulations, see Methods. The results clearly show that the intensity and electric field profiles are completely different in the two cases. For the small ND, the electromagnetic field profile at resonance resembles that of a magnetic dipole, with a clear circulating electric field loop inside the particle. On the other hand, the larger ND shows an electric dipole behaviour, with an antinode in the electric field amplitude at the particle center. It is also clear that the electric field intensity is higher all around the smaller ND ($d=150$\,nm, Figure~\ref{fig:maps}a), and particularly at the ND sides. Although the J-aggregate geometrical distribution around the ND is not well-known, it is natural to assume that a higher field intensity leads to stronger near-field mediated interactions. Furthermore, considering the surface-to-volume ratio as a very rough indication of the ratio between the total electric field energy and the physical space available for the molecules, it is plausible that the smaller ND exhibits a larger splitting. However, given the uncertainty about the molecules position and their dipole orientation distribution, nothing more conclusive than this can be safely stated, and we stress here that particular attention must be devoted when studying dielectric systems with overlapping electric and magnetic dipolar resonances. On the other hand, fewer degrees of freedom can be offered by two-dimensional materials, whose homogeneity and in-plane dipole orientation might result in a simpler system to be studied.\cite{Tserkezis2018,alu2018}
\section{Conclusions}
In summary, we have investigated experimentally and theoretically the light-matter coupling between Mie resonances supported by high-refractive-index nanodisks and excitons in a J-aggregated dye. The lower geometrical symmetry of our nanoparticles compared to the nanosphere case, enables to spectrally separate their MD and ED resonances, and to independently investigate their coupling with the excitonic dipoles. An energy splitting was observed in both cases, larger than the losses in the system, thus setting our system in the strong-coupling regime. Interestingly, we found that the two resonances show a different coupling strength to the excitonic dipoles, as a consequence of their different near-field profiles at resonance.
The coupling strength measured in our system is comparable to those reported for similar metallic systems, thus confirming that high-refractive-index nanoparticles can be a valid alternative platform to plasmonics for room-temperature strong-coupling applications. Moreover, some less lossy high-refractive-index materials may be considered in the future in order to excite even sharper resonances, that can in principle result in a more efficient light-matter coupling and larger splittings. On the other hand, realizing strong light-matter interactions on dielectric nanoparticle systems can be interesting when exploiting the unique features of Mie resonances, such as directional scattering and high intrinsic nonlinearities.

\section{Methods}
\subsection{Sample fabrication}
A silicon-on-sapphire substrate was cleaned in acetone and 2-propanol in an ultrasonic bath for 30 minutes, then dried with a nitrogen steam. Hydrogen silsesquioxane (Dow Corning XR-1541-006) was then spin coated on the sample at 500 rpm for 5 seconds followed by 1200 rpm for 45 seconds, and then baked on an hot plate at 170$^\circ$C for 2 minutes, resulting in a final resist thickness of approximately 200\,nm. The nanostructures were written by electron beam lithography with varying design diameter (in 20\,nm steps) and at different doses, so as to obtain a fine diameter tuning. After developing the hydrogen silsesquioxane in tetramethylammonium hydroxide, the silicon beneath was etched. The etching process is a standard BOSCH process, where after an isotropic etch, a passivation process follows. This sequence is repeated until the desired etch depth is achieved. In our case, since one cycle etches ca. 40\,nm, we etched the sample for 3 cycles ensuring a complete etch of the silicon layer, as clearly evidenced by the change in the substrate appearance from a semi-opaque green colour to completely transparent. Thereafter, a quick 5 minutes BHF bath removed the hydrogen silsesquioxane layer. The sample was finally cleaned in acetone and 2-propanol.
We observe that the electron beam lithography process changes the surface properties of the substrate in the area around the structures, thus making impossible the dye deposition by spin coating without employing surface chemistry. We thus drop-casted the dye from a dilute solution (0.1\,mM methanol solution diluted 1:6 in DI water) depositing a 2\,$\mu$L drop on the nanostructured area and slowly drying it with a gentle nitrogen flow.

\subsection{Optical characterization}
Optical characterization was performed on an inverted optical microscope in dark-field configuration, exciting the sample and collecting back-scattered light through an high numerical aperture (100$\times$, NA$=0.95$) objective lens. The optical image of the sample at the output port of the microscope was collected with a 150\,mm focal length doublet lens and reconstructed on the entrance slits of an imaging spectrometer with a 200\,mm focal length doublet lens. The nanodisks array was carefully aligned with the entrance slits of the spectrometer so as to simultaneously measure 15 particles in a single CCD image. PL spectra from the molecular layer were measured on a Raman microscope, exciting with a diffraction-limited spot at $\lambda=532$\,nm. Low power was used for the excitation, so as to avoid photo-bleaching of the dye.
\subsection{Numerical calculations}
The scattering properties of individual NDs were calculated with an efficient $T$-matrix
method,\cite{waterman1965} namely EBCM,\cite{Mishchenko2002} whose principal characteristic
is that it takes the boundary conditions of continuity of the tangential components of the
fields into account through appropriate surface integrals. The incident field is expanded
into vector spherical waves about the origin, with expansion coefficients $a_{P\ell m}^{0}$,
where $P = E, H$ is the polarization, of electric or magnetic type, and $\ell, m$ are the
usual angular momentum indices. Accordingly, the scattered wave is expanded into vector
spherical waves with expansion coefficients $a_{P\ell m}^{+}$, which are connected to those
of the incident field through the elements of the $T$ matrix, with
\begin{equation}\label{Eq:Tmatrix}
a_{P\ell m}^{+} = \sum_{P' \ell' m'} T_{P\ell m; P' \ell' m'} a_{P' \ell' m'}^{0}~.
\end{equation}
For spherical particles, $T$ is diagonal with respect to $P$ and $\ell$, and independent
of $m$. For non-spherical particles, however, this no longer holds and, strictly speaking,
an unambiguous classification of its eigenvalues in terms of polarization and angular
momentum cannot be made. Nevertheless, for relatively small NDs, one largely predominant
matrix element always exists, and this is what we plot in Figure~\ref{fig:bareSP}c. For
increasing ND size, the contribution and mixing of off-diagonal elements becomes ever more
important, and assigning a predominant character to the resonances -- especially higher-order
ones -- becomes practically impossible. This is strikingly visible in the lower panel of
Figure~\ref{fig:bareSP}c, where the scattering peak at 580\,nm cannot be reproduced by any
scattering matrix element alone, while a higher-order contribution peak (red line) at 550\,nm
does not correspond to a scattering peak due to its inefficient excitation from the incident
plane wave. Scattering and extinction-cross sections are calculated through
\begin{equation}\label{Eq:Crossca}
\sigma_{\mathrm{sc}} = \frac{1}{q^{2}} \sum_{P \ell m}
\left| \sum_{P' \ell' m'} T_{P\ell m; P' \ell' m'} \mathbf{A}_{P' \ell' m'}^{0} \cdot 
\hat{\mathbf{p}} \right|^{2}~,
\end{equation}
\begin{equation}\label{Eq:Crosext}
\sigma_{\mathrm{ext}} = -\frac{1}{q^{2}} \mathrm{Re} \sum_{P \ell m}
\left( \mathbf{A}_{P \ell m}^{0} \cdot \hat{\mathbf{p}} \right)^{\ast}
\sum_{P' \ell' m'} T_{P\ell m; P' \ell' m'} \mathbf{A}_{P' \ell' m'}^{0} \cdot 
\hat{\mathbf{p}}~,
\end{equation}
where $q = \frac{\omega}{c} \sqrt{\varepsilon \mu}$ is the wavenumber of the incident wave of
angular frequency $\omega$ in a homogeneous environment with permittivity $\varepsilon$ and
permeability $\mu$, and $c$ is the velocity of light in vacuum. The matrix elements $\mathbf{A}$
are related to the expansion coefficients $a_{P\ell m}^{0}$ of an incident plane wave with
electric field $\mathbf{E}_{0}$ and polarization $\hat{\mathbf{p}}$ through~\cite{gantzounis2006}
\begin{equation}\label{Eq:MatrixA}
a_{P\ell m}^{0} = \mathbf{A}_{P\ell m}^{0} (\mathbf{q}) \cdot
\hat{\mathbf{p}} E_{0} (\mathbf{q})~.
\end{equation}

To describe the J-aggregate/silicon ND system, we use the Extended Layer-Multiple-Scattering (ELMS) method,~\cite{gantzounis2006} which is ideal for periodic arrays of spherical or cylindrical scatterers. Initially developed for scatterers with spherical symmetry described through Mie theory,~\cite{stefanou1998} the method has been extended through implementation of EBCM, and can efficiently describe transmission and reflection from combinations of stacked arrays of
disks or cylinders of various sizes, with the restriction that all arrays have the same two-dimensional periodicity.~\cite{tserkezis2008,tserkezis2010,tserkezis2011} In our case, we consider ``birthday cakes'' where the role of the cake is played by the silicon ND, and an ``icing'' of thickness 20\,nm and the same diameter is introduced as the molecule. These two-component particles are arranged in square arrays with lattice constant 720\,nm, which ensures that interactions between individual building blocks are minimal, while lattice resonances from the Rayleigh--Wood anomalies are kept outside the wavelength window of interest, as much as possible. Weak interactions for the larger NDs, and small differences between the modelled and real permittivities, explain why slightly smaller particles are required to match the theoretical and experimental spectra, e.g. a diameter of 230\,nm is used in the simulations for the 250\,nm ND of the last panel of Figure~\ref{fig:molecule}. The arrays
are illuminated by p-polarized light, incident at an angle of 70$^{o}$, and reflectance spectra are calculated by integrating the energy flow above the structure over the first Brillouin zone. The coupling strengths are then obtained through the width of the anticrossing in the reflection spectra at zero detuning .The composite arrays are deposited on a semi-infinite sapphire substrate, described by a relative permittivity equal to 3.13. For the permittivity of silicon we use the experimental values of Green,~\cite{green2008} while for the molecule we use a Lorentzian permittivity,
\begin{equation}\label{Eq:Lorentzian}
\varepsilon_{\mathrm{mol}} = \varepsilon_{\infty} - \frac{f \omega_{\mathrm{exc}}^{2}}
{\omega^{2} - \omega_{\mathrm{exc}}^{2} + \mathrm{i} \omega \gamma_{\mathrm{exc}}^{2}}~,
\end{equation}
with $\varepsilon_{\infty} = 2.56$, $\hbar \omega_{\mathrm{exc}} = 2.11$\,eV, $\hbar
\gamma_{\mathrm{exc}} = 0.03$\,eV, and $f =0.35$.

In all EBCM calculations, we truncate the angular-momentum expansions at $\ell_{\mathrm{max}} = 12$. A larger number of matrix elements corresponding to $\ell_{\mathrm{cut}} = 16$ is required at some steps within the method, although they are not involved in the final calculation of cross sections. Integrals at the surface of the particles are calculated with a Gaussian quadrature integration formula, with 4000--6000 integration points, depending on the ND size. The only additional convergence parameter involved in ELMS is the number of reciprocal-lattice vectors involved in plane-wave expansions, and here 161 vectors were required to achieve convergence.

\section{Acknowledgement}
N.~A.~M. is a VILLUM Investigator supported by VILLUM FONDEN (grant No. 16498). The Center for Nano Optics is financially supported by the University of Southern Denmark (SDU 2020 funding). The Center for Nanostructured Graphene is sponsored by the Danish National Research Foundation (Project No. DNRF103). F.T. is grateful to V. Ratano for resilient and lifelong collaboration. C.~T. and C.~W. thank Pano Ramix for stimulating discussions. F.~T. and C.~W. acknowledge funding from MULTIPLY fellowships under the Marie Sk\l{}odowska-Curie COFUND Action (grant agreement No. 713694). 
Simulations were supported by the DeIC National HPC Centre, SDU.

\section{Supporting Informations}
Supporting Information is available upon request to the authors.

\bibliographystyle{unsrt}
\bibliography{references}

\end{document}